# BOSE-EINSTEIN CONDENSATION OF ELECTRONS IN DOUBLY EXCITED STATES OF HELIUMLIKE ATOMS

Kavera V.V.

Krasnodar, Russia

E-mail: vad1809@rambler.ru

**Abstract**. The author suppose a capability of transition doubly excited intrashell configurations of separate atoms to a superconducting state. The conditions of this transition are determined and the experiments for it's detection are offered. The capability of join of superconducting atoms in a superconducting condensate is considered. The capability of spontaneous formation of similar condensates in atmospheres of the Earth and Sun is considered.

## I. INTRODUCTION

It is known, the explanation of superconductivity involves the appearance of electron-electron attraction (its origin and physical nature doesn't matter), which exceeds the Coulomb repulsion under certain conditions.

Author of present work has obtained recently (see [1] and [2]) proofs, that there is an additional not-Coulomb electron - electron attraction together with usual Coulomb repulsion in heliumlike atoms. It is interesting that in the case of intrashell $nl_1nl_2$ doubly excited states, since some value $n$, the additional non-Coulomb attraction will exceed usual electron-electron repulsion.

Though the physical nature of this additional attraction is not clear yet completely, it is possible already now to analyze practical consequences of its existence. The main consequence is a capability of bose-einstein condensation of electrons in separate atoms. It is possible to offer experiments for detection of this effect and it is possible to analyze a capability of observation this effect in a nature. The present work is addition to [2].

## II. CONDITIONS OF TRANSITION OF SEPARATE ATOMS TO SUPERCONDUCTING STATE

In [2] the formula was obtained for additional energy $E_{add}$ in case of doubly excited states $nsns$ ($1S$)

$$E_{add} = -C \cdot \frac{Z}{n} \quad , \tag{1}$$

where $Z$ - charge of the nucleus, $n$ - principal quantum number and $C$ is a constant, the average value of which is close to 1/9. [Hereinafter everywhere in the text Rydberg (Ry) is used as units of energy.]

Making this formula the author used the data of experiment Iemura et al. [3] for case $3s3s$ ($1S$) at $Z = 2$, i.e. for neutral atom of a helium.

But if to use the data of experiment Brotton et al. [4] for $3s3s$ ($1S$), the relation $E_{add}$ by $n$ look as

$$E_{add} = -1{,}5 \cdot C \cdot \frac{Z}{n^{3/2}} \quad , \tag{2}$$

or

$$E_{add} = -\sqrt{2} \cdot C \cdot \frac{Z}{n^{3/2}} \quad . \tag{2'}$$

I.e. relation $E_{add} \sim \dfrac{1}{n^{3/2}}$ occurs instead of relation $E_{add} \sim \dfrac{1}{n}$ .

Let's analyze some theoretical calculations for doubly excited states *nsns* (*1S*).

The values $E$, calculated by some authors and the values of additional energy $E_{add}= E – E_1$ are given in a Tables 1-3, where $E_1$ takes into account all electrostatic interaction in atom (details of calculation $E_1$ by first approximation of a variational method see in [2]). All values are given in Ry.

**Table 1 . The analysis of calculations *Ho* [5]**

| State | -E | -$E_1$ | -$E_{add}$ = - ($E - E_1$) | -$E_{add} \times n^{3/2}$ |
|---|---|---|---|---|
| *2s2s* | 1,55574 | 1,44356 | 0,11217 | 0,31727 |
| *3s3s* | 0,70707 | 0,64305 | 0,06402 | 0,33267 |
| *4s4s* | 0,40199 | 0,36199 | 0,04000 | 0,31998 |
| *5s5s* | 0,25883 | 0,23175 | 0,02708 | 0,30271 |
| *6s6s* | 0,18050 | 0,16097 | 0,01953 | 0,28703 |

**Table 2 . The analysis of calculations *Burgers et al.* [6]**

| State | -E | -$E_1$ | -$E_{add}$ = - ($E - E_1$) | -$E_{add} \times n^{3/2}$ |
|---|---|---|---|---|
| *2s2s* | 1,55574 | 1,44356 | 0,11217 | 0,31727 |
| *3s3s* | 0,70708 | 0,64305 | 0,06402 | 0,33268 |
| *4s4s* | 0,40198 | 0,36199 | 0,03999 | 0,31989 |

**Table 3 . The analysis of calculations *Koyama et al.* [7]**

| State | -E | -$E_1$ | -$E_{add}$ = - ($E - E_1$) | -$E_{add} \times n$ |
|---|---|---|---|---|
| *2s2s* | 1,56017 | 1,44356 | 0,11661 | 0,23321 |
| *3s3s* | 0,72177 | 0,64305 | 0,07872 | 0,23616 |
| *4s4s* | 0,41837 | 0,36199 | 0,05638 | 0,22552 |
| *5s5s* | 0,27657 | 0,23175 | 0,04482 | 0,22409 |

The analysis shows, that two versions of relation $E_{add}$ by $n$ are observed. So for example $E_{add} \sim \frac{1}{n^{3/2}}$ in case of calculations Ho [5] and Burgers et al. [6] and $E_{add} \sim \frac{1}{n}$ in case of calculations Koyama et al. [7].

Let's note, that in the present the experimental data for *nsns* (1S) states are known for $n$ not above 3, i.e. not more than for *3s3s*. For these states Iemura et al. [3] has obtained 0,72063 Ry , that coincides to calculations Koyama et al. [7], and Brotton et al. [4] has obtained 0,70758 Ry, that coincides to calculations Ho [5] and Burgers et al. [6].

The additional precision measurements of *3s3s* (1S) are necessary to select between two versions of relation $E_{add}$ by $n$. It would be better, if the experimenters could obtain additional values at least for *4s4s* (1S).

It is interesting to compare both versions of relation $E_{add}$ by $n$ to energy of elelctron-electron repulsion $E_{rep}$, which can be calculated by integral

$$\iint \psi_1^2 \frac{1}{r_{12}} \psi_2^2 dV_1 dV_2 \quad , \tag{3}$$

where $\psi_1$ and $\psi_2$ - wave functions of separate electrons (see [2]), and $r_{12}$ is the electron-electron distance.

The calculation of integrals (3) gives for *nsns* (*1S*) the asymptotic formula at large $n$

$$E_{rep} = 1,2 \cdot \frac{Z}{n^2} \quad . \tag{4}$$

It is easy to see, that when $n$ grows, the energy of additional attraction $E_{add}$ decreases slower, than energy of a Coulomb repulsion $E_{rep} \sim \frac{1}{n^2}$. It is correctly for both versions, which corresponds to the formulas (1) or (2). I.e. in both cases electron-electron attraction will exceed electron-electron repulsion at some $n$. It is easy to obtain from the formulas (1), (2) and (4), that in case (1) critical $n$ should be not less than 11, and in case (2) critical $n$ should be not less than 52.

The average radius of excited atom will exceed a radius of ground state accordingly in 121 times in the first case and in 2704 times in second case. It is interesting, that in the latter case size of atom (and also average distance between electrons) has the same order, as average size of a Cooper pair in superconductors.

Only experimenters can now answer a question - whether the modern technology allow to obtain necessary $n$ to check up existence bose-einstein condensation of electrons in two-electron atoms or it is impossible in the present ? At least experimenters could a little to increase $n$ values to check up - whether the formula (1) or formula (2) continues to work at large $n$.

### III. PHYSICS OF ADDITIONAL INTERACTION

It is possible to adduce two basic arguments that $E_{add}$ is really additional interaction and that it's physics differs from usual electrostatic interaction.

At first, as was shown in [2], all usual electrostatic interaction in atom is taken into account completely in $E_1$. Main argument is that $E_1$ absolute coincides to experiment for states, in which the overlap of electrons practically is equal to zero (for example for *1snd, 1snf, 1snh* etc. excited states of heliumlike atoms). The deviations from experiment begin only when the overlap of electrons becomes essential (for example for *1sns, 1snp* excited states, *1s1s* ground state and all doubly excited state of a helium, etc.). $E_1$ of third external electron ( in litiumlike atoms) also coincides experiment for states *$1s^2nd$, $1s^2nf$, $1s^2nh$* etc., but in case *$1s^2ns$* и *$1s^2np$* there is also noticeable overlap of an external electron with an internal pair of electrons and nonzero additional energy $E_{add}$. Just the nonzero overlap of electrons creates nonzero nonclassical exchange and correlation effects, which result in occurrence of additional energy $E_{add}$.

Secondly, the analysis shows, that all calculations of heliumlike atoms, which give absolute coincidence with experiment, have one general feature. All of them assume, that the actual atomic configuration results from mixing of various hydrogen-like configurations. This mixing is introduced explicitly - in methods using so-called interaction of configurations, or it is introduced implicitly - by adding additional terms into a hydrogen-like wave function. I.e. actually it is supposed, that the two-electron atom is a dynamic system, in which the transitions happen all the time between various hydrogen-like states. If these transitions were absent, the energy of two-electron atom would coincide with $E_1$. The presence of these transitions results in appearance of additional electron-electron attraction $E_{add}$. Because the transitions between states are accompanied by radiation or absorption of photons, it is possible to present $E_{add}$ as a result of electron-electron interaction through photons, i.e. as a result of exchange by photons. All this very much reminds attraction of electrons in a superconductor as a result of interaction of electrons through exchange by phonons.

Let's note, that in case of ground state 1s1s $E_{add}$ probably is a consequence of exchange by virtual photons, i.e. consequence of zero energy. In case of doubly excited states $E_{add}$ probably is a consequence of exchange by real photons, i.e. consequence of real energy of atomic excitation.

Let's add, that in [2] author has explained some reasons about possible mathematician of an origin of relation $E_{add} \sim \frac{1}{n}$. In addition, it is possible now to notice about version $E_{add} \sim \frac{1}{n^{3/2}}$, what it can be a consequence of relation $E_{add}$ by the first degree of a wave function $\Psi$, which also is proportional $\frac{1}{n^{3/2}}$. It is exotic relation of energy, since usual energy is proportional to the second degree of a wave function .

### IV. REGISTRATION TRANSITION OF SEPARATE ATOMS IN SUPERCONDUCTING STATE

The transition to a superconducting state is possible for registration even in the case of separate atoms. If our suppositions are correct, the atoms can be transformed into an ideal diamagnetic at some critical value *n*. Just the ideal diamagnetism is a main distinctive indication of a superconducting state. The ideal diamagnetism is possible to try to notice even for those atomic configurations, which initially have paramagnetic properties, i.e. which have nonzero moment.

The appearance of large lifetimes at superconducting states (SC-state) of atoms can be one more indication of phase transition in two-electron atom. I.e. these states can be metastable.

## V. INTERACTION OF SUPERCONDUCTING ATOMS

The principle Pauli does not work for electrons, which integrated in boson pairs and therefore it does not prohibit join 2,3 … N superconducting atoms (SC-atoms) in a unified superconducting condensate (SC-condensate). As well as separate SC -atoms, this SC -condensate of excited states can be metastable and it can to have rather large lifetime.

Similar processes could happen spontaneously in nature. It is possible that similar effects could explain at least some of the anomalous plasma-like effects observed sometimes in the atmosphere and in the ionosphere of Earth, such as ball lightning (BL) etc. It is possible, that the SC -atom actually is a «quant» of BL, which is a SC-condensate of excited states. Then the obtaining of SC-atom in laboratory will mean obtaining by first artificial BL.

Similar processes could also help to explain some until now misunderstood processes in atmosphere of the Sun and in other cases of observation of plasma in a nature. For example similar structures could explain process of accumulation energy and it's explosive release in solar flares. Or for example condensation of SC –atoms with a nonzero moment (orbital or - and spin), could explain spontaneous origin of vortical structures, spontaneous origin of magnetic fields in various cases etc.

Let's note, the transition to a SC-state is possible for atoms of any chemical elements, but may be it is most simply in case of two-electron atoms, especially in case of neutral atom of a helium. This element is in small quantity in atmosphere of the Earth and it is widespread very much in atmosphere of the Sun. The negative ion of hydrogen also is widespread in atmosphere of the Sun and in atmosphere of the Earth.

## VI. CONCLUSION

The author of the present work hopes that the relations obtained by him, as well as practical conclusions and the predictions of new physical effects can interest theorists and experimenters working in the field of physics of atom and molecules, physics of condensed matter and physics of plasma.